# Science Tree: A Platform for Exploring the Brazilian Academic Genealogy

João M. M. C. Cota, Alberto H. F. Laender, Raquel O. Prates
Department of Computer Science
Universidade Federal de Minas Gerais
31270-901, Belo Horizonte, MG
Brazil

## ABSTRACT

Identifying and studying the formation of researchers over the years is a challenging task, as the current repositories of theses and dissertations are cataloged in a decentralized manner in different digital libraries, many of them with limited scope. In this paper, we take a step forward towards building a large repository to record the Brazilian academic genealogy. For this, we collected data from the Lattes platform, an internationally recognized initiative that provides a repository of researchers' curricula maintained by the Brazilian National Council for Scientific and Technological Development (CNPq), and developed a user-oriented platform to generate the academic genealogy trees of Brazilian researchers from them, also providing additional data resulting from a series of analyses regarding the main properties of such trees. Our effort has identified interesting aspects related to the academic career of the Brazilian researchers, which highlight the importance of generating and cataloging their academic genealogy trees.

## 1 INTRODUCTION

Throughout history, many researchers have made remarkable contributions to science, not only advancing knowledge, but also forming new scientists. A relevant way to characterize the impact of a researcher is by means of her/his academic genealogy [2, 9, 20, 25, 31, 32]. The term academic or scientific genealogy refers to the organization of a tree of scientists and scholars structured according to the academic supervision relationships (master and doctorate) carried out by them throughout their respective careers [21]. In this context, a researcher's academic genealogy can impact even the results of a specific research. For example, Hirshman et al. [20], when analyzing the medical literature related to the glioma surgery[1], concluded that it has been greatly influenced by medical academic genealogies, i.e., the authors of this literature are usually part of a medical team whose members have been mentored over the years by other members from the same team.

However, identifying and studying the formation of researchers over the years is a challenging task, as the current repositories of theses and dissertations are cataloged in a decentralized manner in different digital libraries, many of them with limited scope regarding the research area they address. Thus, despite the many works that seek to analyze the dynamics of scientific production in terms of collaboration networks [1, 3, 7, 12, 26, 27], there are still few efforts aiming to study and preserve the academic supervising process. Among the pioneers, Jackson [21] reports the first efforts made to build the Mathematics geneology tree, which gave rise to the *Mathematics Genealogy Project*[2]. Initially, the project had the collaboration of more than 3,500 researchers who sent their data by personal correspondence, currently counting with more than 256,000 names, which highlights the relevance of integrating, on a single platform, data related to academic supervision from different knowledge areas, as corroborated by some pionnering initiatives [6, 19, 25, 32].

After a preliminary effort to build academic genealogy trees using data extracted from the Networked Digital Library of Theses and Dissertations (NDLTD) [13], another work from our research group [14] presented a first study on the Brazilian academic genealogy, based on data collected from the Lattes platform[3]. Maintained by the Brazilian National Council for Scientific and Technological Development (CNPq), this platform is an internationally recognized initiative [22] that provides a repository of researchers' curricula vitae and research groups, all integrated into a single system. Since all researchers in Brazil (from junior to senior) are required to keep their curricula updated in this platform for being entitled to apply to any grant, it provides a great amount of information about their research activities and scientific production, including all their master and doctoral supervision activities. Based on such previous experiences, we started an effort to construct a specific platform aimed at exploring the Brazilian academic genealogy. Named *Science Tree*[4], this new platform not only allows the visualization of the academic genealogy tree of all researchers associated with an academic or research institution in Brazil, but also makes it possible to interact with such trees, thus providing its users with a clear view of how they have evolved and, more specifically, how such researchers have contributed to consolidating specific research areas. For this, we crawled the entire Lattes platform and collected the curricula of all researchers holding a PhD degree. Next, we developed a framework to extract specific data from the collected curricula and properly identify (disambiguate) the respective researchers using up-to-date techniques [17], in order to establish their supervisor-supervisee relationships and build their respective academic genealogy trees. We also conducted two specific user evaluations of the developed platform: a broader one based on a questionnaire answered by 286 researchers from distinct Brazilian academic and research institutions, and an in-depth evaluation, conducted with a small number of selected prospective users in a controlled environment, which provided us with many insights for improving some of its features [8].

---

[1]Glioma is a type of tumor that starts in the glial cells of the brain or the spine (see https://en.wikipedia.org/wiki/Glioma, accessed in February 2021.

[2]https://www.genealogy.math.ndsu.nodak.edu, accessed in February 2021.
[3]http://lattes.cnpq.br, accessed in February 2021.
[4]*Árvore da Ciência* in its Portuguese version and *Árbol de la Ciencia* in the Spanish one. Available at https://sciencetree.net, accessed in February 2021.

Thus, the rest of this paper is organized as follows. Firstly, Section 2 presents a brief overview of similar initiatives, whereas Section 3 provides a functional view of the *Science Tree* platform. Next, Section 4 describes a case study highlighting the main features of the platform and Section 5 summarizes the results of the two user evaluations conducted with it. Finally, Section 6 presents our conclusions and highlights some directions for future work.

## 2 SIMILAR INITIATIVES

In recent years, efforts have been made by some research groups to better understand the process of formation of new researchers in distinct knowledge areas by exploring their academic genalogy [2, 9, 12–14, 21]. Despite that, there are still few efforts to preserve the academic genealogy of specific scientific communities or even countries. Thus, in this section we present an overview of five existing initiatives that aim at preserving the academic genealogy of a specific knowledge area or scientific community. In doing so, we intend to contrast the facilities provided by the current version of the *Science Tree* platform with those of these existing initiatives.

The five initiatives chosen to address here are the *Mathematics Genealogy Project*[5], *The Academic Family Tree*[6], the *American Meteorological Society (AMS) Academic Family Tree*[7], the Acacia Platform[8] and *The Gold Tree*[9]. These platforms were chosen because they are today the initiatives that cover the largest number of researchers and also due to their relevance to the area or scientific community they address. In what follows, we briefly describe each one of these initiatives in order to overview their main facilities and then provide a means to compare them with our proposed platform.

***Mathematics Genealogy Project.*** This is one of the first initiatives aimed to provide a platform for visualizing the academic genealogy trees of a specific scientific community [21]. As described on its website, this platform provides data on mathematicians associated with academic institutions from all over the world. It also enables institutions that carry out research in the area of mathematics to register their researchers so that they can have their data widely disseminated. Currently, the platform stores data from more than 253,000 mathematicians around the world, which are presented in textual form. Searches on this platform can be carried out by simply using the name of a specific mathematician or in a more refined way by means of its advanced search features. The data available about a researcher are name, obtained degree, affiliated institution, title of her/his dissertation or thesis, name(s) of her/his supervisor(s) and the list of her/his former MSc/PhD students. The platform also provides two specific metrics to express the mentoring performance of a researcher: *number of supervisees* and *number of academic descendents*.

***The Academic Family Tree.*** This platform originated from another similar initiative, the *Neurotree*[10]. Its current version is an expansion of that original platform, which started in January 2005 with the aim of documenting the academic supervisions in the Neuroscience area [10] and presenting them in the form of a genealogy tree. Right after the creation of the *Neurotree*, its maintainers realized that the Neuroscience had relationships with other scientific areas and such relationships could be as important as those of the Neuroscience itself, leading to the incorporation of new areas into the platform that was then renamed *The Academic Family Tree*. The homepage of this platform lists all websites that make part of it, grouped by affinity of their respective knowledge areas. On that page, it is possible to carry out a general search on the entire platform or to browse a certain knowledge area to find a specific researcher associated to it. It also provides 15 metrics that allow its users to assess the relevance of the existing trees. The platform lists today more than 748,000 researchers with more than 687,000 connections between them. As a result of the search for a researcher, her/his tree is displayed, showing not only her/his academic descendents, but also her/his academic ancestors (direct supervisors and their respective supervisors). From the selected researcher, it is possible to perform interactions using specific commands that allow one to expand the nodes horizontally and navigate vertically through them, showing their direct descendants and the descendants of them. For each researcher it is also shown the name of her/his current institution.

***AMS Academic Family Tree.*** Another interesting initiative is provided by the Florida State University that hosts the website of the academic genealogy tree of American meteorologists. Maintained by the American Meteorological Society[11], this website makes it possible to explore the trees of different communities of American meteorologists [19]. In addition, a researcher may also request her/his inclusion in the tree by means of a specific function (*Add Yourself/Someone Else*). However, the website does not offer an automatic way of generating or viewing the trees, so they are compiled from time to time and made available in image format. It also does not provide any metrics to assess the trees. Currently, the website includes the academic genealogy tree of more than 6,000 American meteorologists.

***Acacia[12] Platform.*** Launched in April 2018, this platform aims to register the formal supervising relations in the context of the Brazilian graduate programs. By also using data collected from the CNPq's Lattes platform, it generates academic genealogy graphs in which each vertex represents a researcher and each edge corresponds to a concluded graduate supervision (MSc or PhD) [9]. Maintained by a research group from the Federal University of ABC, in the state of São Paulo, this platform focuses mainly on the researchers' academic education and supervising activities, providing a table-oriented representation of their concluded MSc and PhD supervisions. Another point that should be noted is that it provides a simple navigation scheme involving the researcher's descendants and ancestors, but without any graphical representation of these supervising relationships. On the other hand, it provides seven specific metrics on the researcher's academic genealogy, which express relevant information about their career. These metrics are: *Descendency* (number of direct and indirect descendants), *Genealogical Index* (the largest number $g$ of descendants who have the same number of descendants), *Fecundity* (number of direct descendants), *Fertility* (number of direct descendants who have at least one descendant), *Generations* (number of generations of a researcher expressed by the height of her/his tree), *Relationships* (number of supervision relations in the academic descendancy of a researcher) and *Cousins* (number of supervisees who have a common academic "grandparent" but distinct supervisors). All these metrics are compiled based on data extracted from the researchers' curricula. These metrics can also be calculated by grouping the researchers according to the CNPq's knowledge area classification scheme [11]. The *Acácia Platform* lists today more than 1.2 million researchers involved in more than 1.4 million supervising relations.

---

[5] https://www.genealogy.math.ndsu.nodak.edu, accessed in February 2021.
[6] https://academictree.org., accessed in February 2021.
[7] http://moe.met.fsu.edu/familytree, accessed in February 2021.
[8] http://plataforma-acacia.org, accessed in February 2021.
[9] http://thegoldtree.c3.furg.br, accessed in February 2021.
[10] http://neurotree.org, accessed in February 2021.
[11] http://www.ametsoc.org, accessed in February 2021.
[12] Acacia is a genus of plants that groups together numerous species of shrubs and trees known by the common name of acacias.

| Platform | Genealogy Representation (Tabular or Graphical) | Data Visualization (Textual or Graphical) | Number of Available Metrics | Focus of the Platform |
|---|---|---|---|---|
| Mathematics Genealogy Project | Tabular | Textual | 2 | Worldwide |
| The Academic Family Tree | Both | Both | 15 | Worldwide |
| AMS Academic Family Tree | Tabular | Textual | None | Single Country |
| Acacia Platform | Tabular | Textual | 7 | Single Country |
| The Gold Tree | Graphical | Graphical | None | Single Country |

**Table 1: Major characteristics of the five similar initiatives.**

***The Gold Tree.*** This platform provides an environment for visualizing the academic genealogy trees of Brazilian researchers who have their curricula registered in the CNPq Lattes platform [24]. According to its website, The Gold Tree platform relies on data from more than 570 thousand MSc dissertations and PhD theses currently registered in the Lattes platform to provide its users with the experience of visualizing and exploring the academic genealogy trees derived from them. Developed by the Information Management Research Group from the Federal University of Rio Grande, in the state of Rio Grande do Sul, this platform allows its users to explore the academic genealogy of a large number Brazilian researchers. However, it does not offer any metrics or additional data to provide its users with a more effective view of the actual role of the researchers as academic supervisors.

***Final Comments.*** Table 1 summarizes the main characteristics of the five platforms addressed in this section. As we can see, the genealogy representation of these platforms is predominantly tabular, with *The Academic Family Tree* and The Gold Tree providing a graphical view of the trees. Notice that the representation adopted by the platforms has a direct impact on how the trees are visualized, which makes the *The Academic Family Tree* the most versatile one in this regard. With respect to the metrics, *The Academic Family Tree* platform is also the one that offers more options (15), thus providing its users with more information when analyzing the trees, followed by the Acacia Platform with seven and the Mathematics Genealogy Project with only two. The AMS Academic Family Tree and The Gold Tree provide no metrics. However, the five platforms have in common limitations regarding how their users interact with them. The focus of each platform also varies, with the *Acacia* and *The Gold Tree* platforms dealing with data from a specific national repository, in this case the CNPq Lattes platform, while the *Mathematics Genealogy Project* and *The Academic Family Tree* record data from institutions in different countries around the world to compose their repository, being generally fed by the researchers themselves or by third parties, for example, institutions interested in maintaining the trees of their faculty members or researchers. Finally, updates to the *AMS Academic Family Tree* are sent by the researchers themselves using an institutional email, and then manually uploaded to the platform by its maintainers.

## 3 FUNCTIONAL VIEW OF THE PLATFORM

Figure 1 shows the major steps performed by the *Science Tree* platform for generating the academic genealogy tree of a Brazilian researcher based on data extracted from her/his curriculum vitae available on the Lattes platform. As we can see, after a user submits a researcher's name through the initial query page of the *Science Tree* platform, a search is carried out on the repository of curricula vitae collected from the Lattes platform to retrieve the required data for building the respective academic genealogy tree. As the data are retrieved, the researcher's tree and its respective metrics are generated, and then displayed to the user together with specific data about the researcher and her/his tree, such as name, academic background and institution, metric values expressing different aspects of her/his academic genealogy and a timeline of her/his concluded supervisions as recorded on the Lattes platform. It is worth mentioning that all such data are kept in local repositories maintained by the *Science Tree* platform itself and made publicly available for access by its users on the World Wide Web.

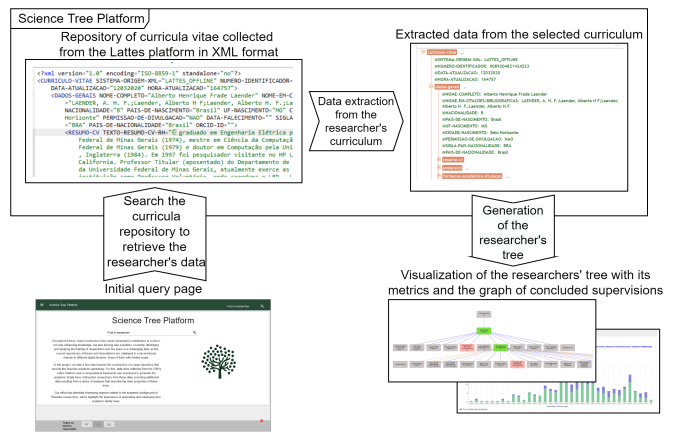

**Figure 1: Data flow on the *Science Tree* platform.**

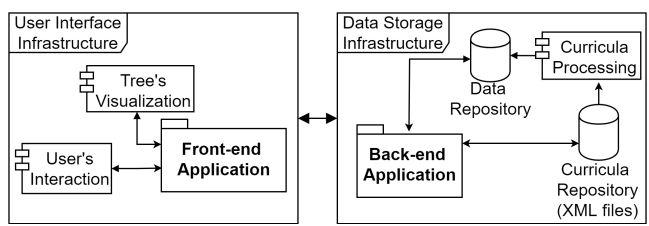

Figure 2: Architecture of the *Science Tree* platform.

As shown in Figure 2, the platform consists of two main applications: the *Front-end* and the *Back-end*. The *Front-end Application* establishes the interface with the users, processing their queries and structuring the data they will be able to access while using the platform, for example, when they navigate through the academic genealogy trees visualizing the data of the respective researchers or when they want to have access to the values of certain metrics that express the structure of a specific tree, thus characterizing the academic performance of a researcher. On the other hand, the *Back-end Application* intercepts requests the *Front-end Application* and searches the platform's Data Repository in order to calculate, for example, the values of the genealogical metrics provided for a researcher, and then generate her/his chart of concluded supervisions, as we will illustrate in the case study presented in Section 4.

After the relevant data are extracted from the collected curricula and stored in the data repository, they are directly consumed by the *Back-end Application* that interprets the queries performed by the user through the *Front-end Application*. This is how a user interacts with the platform in order to visualize the academic genealogy tree of a specific researcher. The *Back-end Application* is also able to directly access the Curricula Repository to provide additional data about the researcher whose genealogy tree is being explored by a user, such as her/his short resume available in the Lattes platform.

The user interface of the *Science Tree* platform was conceived based on a minimalist design, with few visual elements and simple textual descriptions. These characteristics help its users to better understand the data presented to them, in addition to reducing the navigation time when exploring the trees. Besides, since the platform provides several metrics that allow its users to quantify different aspects of the researchers' academic genealogy, it was important that its interface would also allow them to obtain a more analytical view of the performance of such researchers with respect to the role they play as supervisors.

To provide this, the interface was implemented using the visualization libraries *ReactJs*[13] and *VisJs*[14], both written in JavaScript. These libraries were chosen due to their capability to capture the users' interactions and build sophisticated graphic interfaces, thus supporting not only the visualization of the academic genealogy trees, but also allowing the users to interact with them without re-rendering the entire page. The access to the platform is done by using any electronic device (computer, smartphone or tablet) connected to the Internet and having a browser. The platform is hosted on the site https://sciencetree.net and does not require any prior registration of its users for accessing it.

---

[13]https://reactjs.org, accessed in February 2021.
[14]https://visjs.org, accessed in February 2021.

As we can see from Figure 2, the generation of the researchers' academic genealogy trees requires preprocessing their curricula in XML format, previously collected from the Lattes platform. For this, we use a *Python* implementation of the algorithm proposed by Dores et al. [14]. The resulting supervisor-supervisee relationships are then stored in the Data Repository as a directed graph, so that they can be visualized in the form of academic genealogy trees and have the values of their respective metrics generated.

To make this data available quickly and securely, a Web API was developed using the C# programming language along with the framework .NET Core [18]. For this, a graphical user interface was provided for visualizing and exploring the generated trees, using the *ReactJs* library implemented in *JavaScript*, which allows the creation of unique pages with dynamic content that are directly rendered in the user's browser. For the generation of the trees, we used the *VisJs* library due to its graphic facilities such as *zoom*, clicks on nodes and dynamic expansion of graphical images without reloading the respective pages, which is an important resource to enable the expansion of the researchers' trees to provide the visualization of their respective descendants. To properly support the proposed platform, it was necessary to create a specific computing environment to meet its characteristics. The implementation of this environment was carried out by using the *Docker* platform[15], thus facilitating the allocation of additional resources, such as the *MariaDB* DBMS[16], adopted for storing the data extracted from the Lattes curricula, and specific tools for handling XML documents.

## 4 CASE STUDY

To illustrate the major features of the *Science Tree* platform, we now present a brief case study based on the academic genealogy tree of Prof. Crodowaldo Pavan[17] (1919-2009), one of the most renowned Brazilian geniticists. Graduated in 1941 in Natural History by the Faculty of Philosophy, Sciences and Letters of the University of São Paulo (USP), he obtained his Doctorate degree, also from USP, in 1944 in the area of Genetics for his studies on the environmental adaptation and evolution of the blind fishs from the Iporanga caves, located in the state of São Paulo, Brazil. He is also recognized for his studies on the cytogenetics of the American Rhynchosciara fly, known for its giant chromosomes, which contributed to demonstrate that the structure of genes and chromosomes are modified by infections. In addition to his academic contributions at USP, Prof. Pavan also played important admistrative roles in his carrer as, for instance, Chairman of CNPq, the Brazilian National Council for Scientific and Technological Development, from 1986 to 1990.

Prof. Pavan's curriculum was included in the Lattes platform in 2008, shortly before his death in 2009. However, since the goal was just to create an entry in his honor in that platform, the submitted curriculum included just basic personal data and registered a single

---

[15]https://www.docker.com, accessed in November 2020.
[16]https://mariadb.org/, accessed in November 2020.
[17]https://en.wikipedia.org/wiki/Crodowaldo_Pavan, accessed in February 2021.

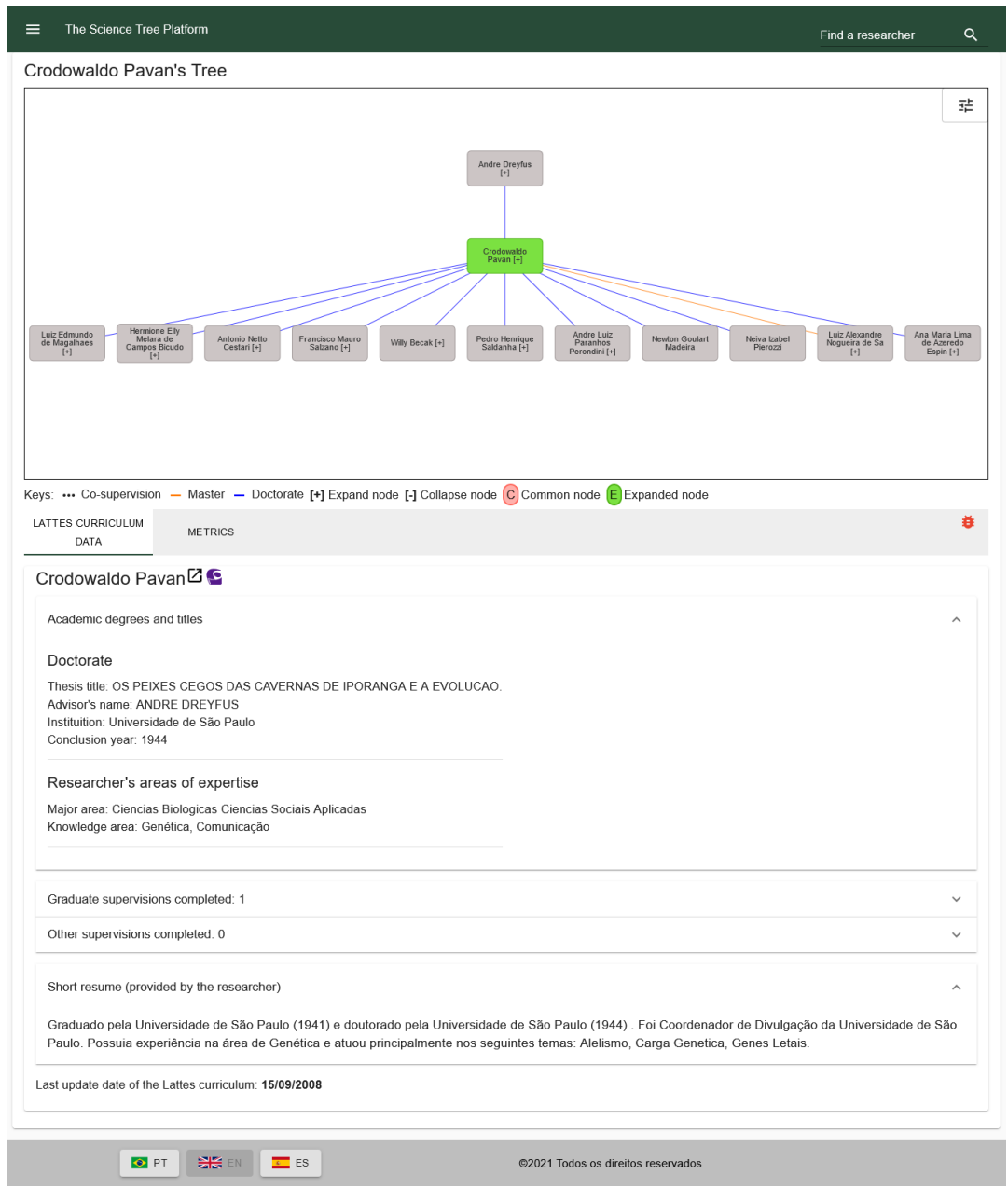

**Figure 3: Prof. Crodowaldo Pavan's page on the *Science Tree* platform showing the first level of his academic genealogy tree and some data related to his carrer.**

Doctorate supervision. Nevertheless, as it will be shown next, the academic genealogy tree of Prof. Pavan on the *Science Tree* platform includes 6,322 direct and indirect descendants. For its generation, not only data from his own curriculum vitae were used, but also from 10 other ones found on the Lattes platform, whose researchers indicated Prof. Pavan as their supervisor, thus summing up 11 supervisions, of which 10 were identified by the algorithm proposed by Dores et al. [14] for generating academic genealogy trees based on data extracted from researchers' curricula available on the Lattes platform.

Figure 3 shows the English version[18] of Prof. Pavan's page on the *Science Tree* platform, which displays the first level of his tree, with the nodes of his 11 direct academic descendants and the contents of his LATTES CURRICULUM DATA tab. This tab displays his name, followed by the icon  that allows accessing the same content in a new browser tab and the Lattes platform icon  that redirects the user to Prof. Pavan's page available on that platform. Next, it is possible to see that his curriculum registers a single graduate supervision and no other forms of supervision. Prof. Pavan's personal data shown in the Academic Degrees and Titles section are the title of his thesis, the name of his supervisor, the institution and year in which he defended his thesis, and his respective major areas and areas of expertise as classified according to the CNPq knowledge classification scheme [11]. Next, it is possible to see that his curriculum registers a single graduate supervision and no other forms of supervision. Finally, there is a sequence of four expandable sections that, when expanded, shows a summary of specific data from his curriculum as available on the Lattes platform.

---

[18]The *Science Tree* platform is trilingual. Thus, although most of its content extracted from the Lattes platform is in Portuguese, all its pages can be displayed in Portuguese, English or Spanish, thus facilitating the users' navigation through the academic genealogy trees of the Brazilian researchers. For this, the user just needs to press the respective button (PT, EN or ES) on the bottom of any page.

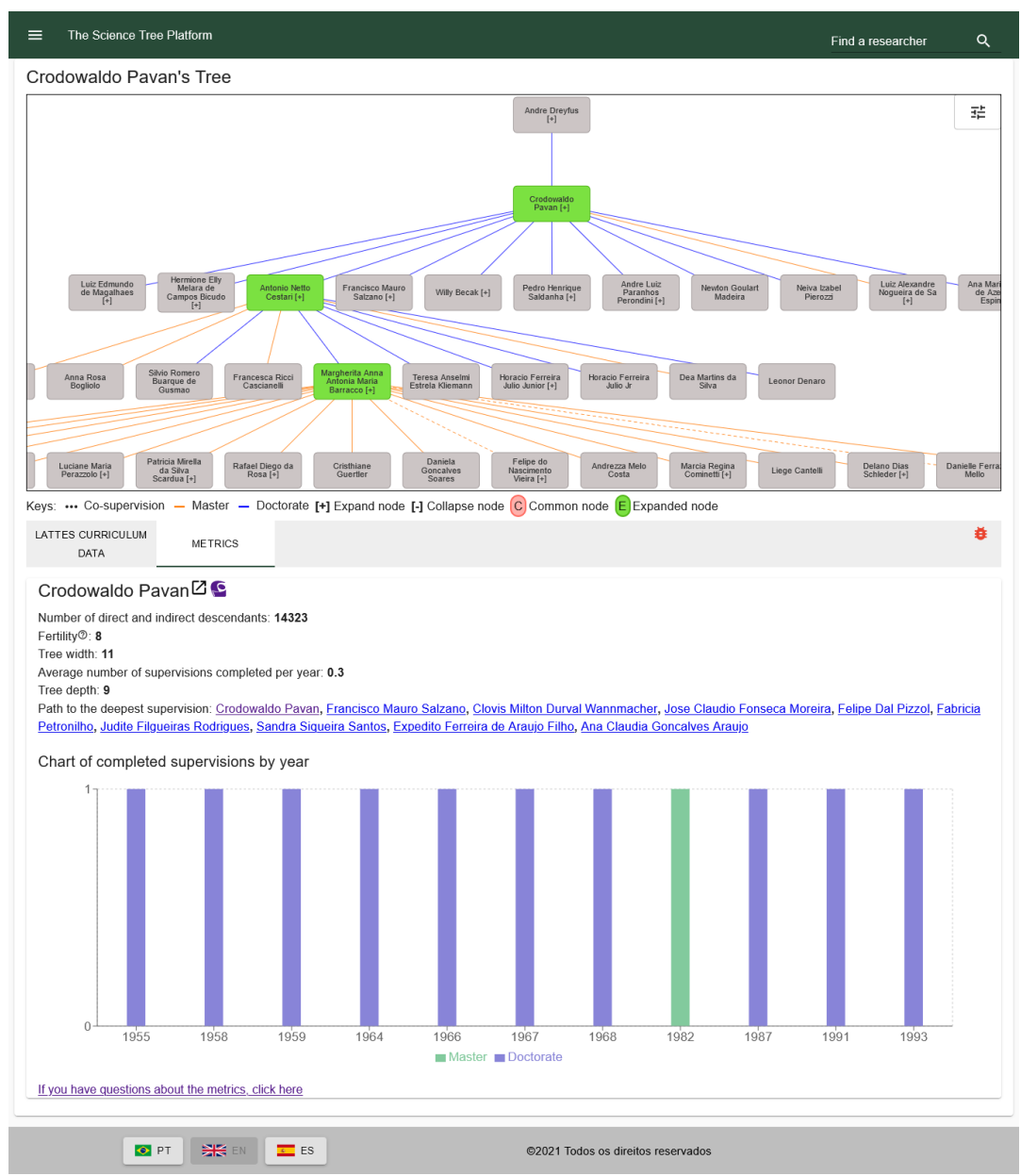

**Figure 4: Prof. Crodowaldo Pavan's tree expanded to its third level for showing a specific path of its academic lineage (green nodes). Below a specific tab shows some metrics related to his tree and his chart of concluded supervisions by year. In the trees, blue lines mean doctoral supervisions and orange lines master's ones.**

Figure 4 shows another instance of Prof. Pavan's page on the *Science Tree* platform that displays a more detailed version of his academic genealogy tree with some of its nodes expanded (those in green) and his metric tab open just below. Examining this other page in more detail, we can see that, after the expansion of the node of his academic descendant Antonio Netto Cestari, it was possible to visualize another generation of indirect descendants of Prof. Pavan, like the researcher Margherita Anna Antonia Maria Barracco, whose expanded node allowed the visualization of another generation of Prof. Pavan's academic descendents.

That page also shows some important metrics that express a researcher's mentoring effort in terms of her/his academic genealogy tree, among them *tree depth* (length of the path from the root to the deepest node), *fertility* (number of direct descendants who have completed at least one supervision), *tree width* (number of direct descendants) and *average number of concluded supevisions per year*. In the case of Prof. Pavan's academic trajectory, the values of these metrics are 8, 9, 11 and 0.3, respectively, the latter being calculated as the width of his tree divided by the difference between the year of his last supervision and the year of his first supervision, that is, $\frac{11}{1993-1955} = 0.3$. Just below the metric values, it is also possible to see the path to the deepest supervision in his tree.

Finally, the chart of concluded supervisions per year is a complementary way of expressing a researcher's mentorship effort by showing, by means of a bar chart, in which blue bars represent doctoral supervisions and green bars master's supervisions, the total number of supervisions completed at each specific year. In the case of Prof. Pavan, this graph indicates, based on data extracted from the Lattes platform, that he supervised seven doctoral students that concluded between 1955 and 1968, one master's student that concluded in 1982 and three doctoral students that concluded between 1987 and 1993. However, the great impact of Prof. Pavan in mentoring new researchers can be measured by his number of descendants, which sums up 6,316, thus meaning that his 11 supervisees contributed to the formation of an expressive number of masters and doctors throughout their academic careers.

## 5 EVALUATION BY END USERS

The overall process of software development usually involves several distinct steps, such as requirements analysis, design, implementation, testing and production [29]. Among the different types of software testing, there are those that aim to analyze the users' perspective with respect to a developed system or application, as well as those that evaluate its functionality and overall design. In the case of the *Science Tree* platform, two distinct user evaluations have been carried out with prospective users to collect data related to both their perception of the usefulness of the proposed platform and the easiness of interacting with it, in order to identify existing difficulties and provide suggestions for future improvements [15, 30].

As suggested by Prates and Barbosa [28], the evaluation of an application with real users is important to collect insights from its target audience with respect to the facilities provided by it. Thus, in the case of the *Science Tree* platform, an evaluation with representative end users of its target audience would allow us to collect relevant indicators regarding the facilities it provides, as well as to identify interaction aspects that could be improved. Several techniques might be used for evaluating the quality of a system from the perspective of its end users. The selection of such techniques must take into account the objective to be achieved with the proposed evaluation [4]. In this work, we applied two specific user evaluation techniques: a quantitative one based on a questionnaire and a qualitative one carried out with a small group of previously selected prospective end users and conducted in a controlled environment.

The evaluation based on questionnaire aimed to carry out a broader data collection, mainly focused on obtaining quantitative indicators of aspects related to the platform's usability. It also included an optional open-ended field in which respondents could include a comment or suggestion. The end user evaluation, on the other hand, aimed to obtain in-depth information about the participants' views of the proposed platform and also regarding their experience when using it. Thus, considering that these two techniques are complementary, they would allow not only a broad analysis of the users' perspective with respect to the *Science Tree* platform, but also the identification of those aspects that could be improved in its interface, as well as in its interaction mode [16, 23]. The remaining of this section is organized in two subsections. In the first one, we present the methodology adopted in each evaluation and describe its application, whereas in the second one we present the results of our analyses.

### 5.1 Methodology

***Evaluation Based on Questionnaire.*** For this evaluation, we used a questionnaire specially designed for it by using the *Google Forms*[19] service. More specifically, the questionnaire was divided into three distinct parts. The first one, mandatory, was aimed to identify the respondents' profile and consisted of five questions meant to anonymously collect the following data about them: gender, age, area of expertise, level of education and occupation. The second part included 13 multiple-choice questions based on the Likert scale[20] for assessing the users' impression about the platform, considering the following six alternatives: "Strongly Agree", "Agree", "Neutral", "Disagree", "Strongly Disagree" and "I have no opinion". The statements that comprised this part of the questionnaire were:

(1) The visualization of the researchers' metrics is clear;
(2) The metrics on the researcher's performance as supervisor are useful;
(3) When selecting a researcher, the data presented about her/him is clear.
(4) The platform is aesthetically pleasant;
(5) The trees are clearly visualized;
(6) The colors used on the platform do not cause confusion;
(7) The filter options available are clear;
(8) The platform is easy to use;
(9) It was easy to learn how to use the platform;
(10) Navigating through the platform is intuitive;
(11) The platform is useful for you as a researcher;
(12) Exploring the academic genealogy trees is useful to assess the researchers' performance;

---

[19]https://docs.google.com, accessed in February 2021.
[20]Proposed by Rensis Likert (1903-1981), Professor of Psychology and Sociology, and Director of the Michigan Institute of Psychology for many years, this is one of the scales most used for opinion polls.

  (13) The visualization of the advisor-advisee relationships by an academic genealogy tree is useful.

It is important to notice that such statements were elaborated considering a logical sequence among them, but they have been shuffled and then randomly inserted in the questionnaire presented to the researchers to avoid any bias when being answered by them.

Finally, the third part of the questionnaire included a single three-choice ("Yes", "No" and "In some points") question to assess whether the user had any difficulty to understand the platform data flow, followed by an open-ended question in which the respondents could optionally include any suggestions, comments or criticism.

Regarding its application, the questionnaire was sent to a list of 5,834 *e-mails* of researchers associated with the National Institutes of Science and Technology (INCTs), a program sponsored by CNPq to support multidisciplinary research initiatives[21] that involved several Brazilian research institutions. All together, 291 of the contacted researchers answered the questionnaire, 286 of whom were considered for having their responses analyzed, because they completed the questionnaire in its entirety, corresponding to 4.9% of the researchers contacted by *e-mail*. Regarding the final open-ended question, 147 out of the 286 respondents (51,4%) left a comment. However, 15 of them were not actual comments, but just included an e-mail address or a congratulatory message left by the respondent. Thus, at the end, 132 comments (corresponding to 46,2% of the respondents) were considered valid and addressed in our qualitative analysis (see Subsection 5.2).

Out of the 286 respondents, 179 (62.5%) declared themselves male, 103 (36.0%) female, and 4 (1.5%) chose not to declare their gender. The age of the respondents was spread over a spectrum ranging from 24 to 75 years, with the highest number of respondents in the age group from 50 and 59 years. This fact may indicate that most respondents are researchers still active and, eventually, with a consolidated career in their respective research areas. Another interesting fact is the existence of 19 respondents (6.6%) over 70 years old, showing that such researchers, possibly retired, are also interested in knowing their own academic genealogy trees and, possibly, exploring the trees of others colleagues. With respect to their professional occupation, 220 (76.9%) respondents were university faculty members, 54 (18.9%) researchers from public research centers, and 6 (2.1%) lab technicians. The remaining included six students (2.1%). Finally, the group of respondents included researchers from distinct knowledge areas such as Biochemistry, Biology, Chemistry, Computer Science, Engineering, Health, Mathematics, Pharmacology, Physics, and Social Sciences, among others.

***Evaluation with Prospective Users.*** For this second evaluation, we selected as participants faculty members from relevant graduate programs at the Universidade Federal de Minas Gerais (UFMG) and also from other federal institutions in the region, as the participation required a face-to-face session. Thus, we sent invitations to 27 faculty members from distinct knowledge areas and in different levels of their academic career (assistant, associated and full professors). The invitation included a brief explanation of the evaluation process and a short description of the *Science Tree* platform. From the total of 27 invitations sent out, we obtained seven positive replies. The evaluation process involved performing a few tasks using the the *Science Tree* platform and participating in a semi-structured interview on their experience and opinion about the platform.

Only one participant was present at each evaluation session. During such sessions, the participants were requested to perform three predefined tasks that involved identifying specific information on a researcher's genealogy tree or associated with the metrics presented. Next, they were asked to freely explore a tree of their interest (their own or of a colleague) and to think aloud [23] as they did. Afterwards, a semi-structured interview was conducted that explored the following questions:

  (1) What was your general opinion about the platform? Please, cite its main strengths and weaknesses.
  (2) Did you find any difficulty in performing the given tasks? If yes, which ones and why?
  (3) About the platform metrics chosen in some of the tasks, which ones did you find interesting and why?
  (4) What did you think about the visualization of the trees? Justify.
  (5) What did you find about the configuration of the visualization options?
  (6) Would you use the platform in your day-to-day activities? In which circumstances?
  (7) Do you have any suggestion to improve the platform?

The user evaluations took place between February and March 2020 in the dependencies of the Universidade Federal de Minas Gerais (UFMG). Out of the seven participants in this evaluation, aged between 30 and 60 years, six were from the Universidade Federal de Minas Gerais and one from the Universidade Federal de Ouro Preto. Six of them were full professors with an average of 20 concluded supervisions in their respective graduate programs, and one was an assistant professor and had not supervised any graduate students yet. Their research interests were related to the areas of Electrical Engineering, Mathematics, Physics, Sociology, Biology and Communication. The sessions took on average 37 minutes. The users' interactions and comments were recorded, and later transcribed for the analysis.

### 5.2 Results

We next present the results of of the quantitative and qualitative analyses conducted on the collected data. The quantitative analysis is based on Likert scales statements of the survey, whereas the qualitative analysis presents the overall results obtained from both the comments left in the open-ended question of the questionnaire and the users' evaluation.

***Quantitative Analysis.*** The analysis presented in this subsection is based on the 13 multiple-choice questions of the questionnaire. The graphs in Figure 5 show the distribution of the respondents' opinions for each one of these questions, grouped by the following aspects related to the user evaluation conducted with the *Science Tree* platform: *metrics and data presentation*, *visual aspects*, *ease of use*, and *usefulness*.

As we can see, regarding the graphs on *metrics and data presentation* shown in Figure 4(a), the majority of the respondents (over 63.0%) had positive attitudes about the visualization and usefulness of the researchers' metrics, the same happening with respect to the data presentation. Considering now the graphs on *visual aspects* of the platform shown in Figure 4(b), the majority of the respondents (over 56.0%) also agreed with the statements of the respective questions. However, when asked about the filter options[22] being clear, the users' agreement decreased to 44.8%, with 28.7% of respondents disagreeing with the statement and 26.6% being neutral or not having an opinion. This indicated to us that the filter button was not clearly perceived by the users, and that it would be worth exploring whether the problem was its location (it is now located on the right top corner of the platform's main page, as shown on the example in Figure 3) or the icon chosen to represent it in order to fix the problem. With respect to *ease of use* (Figure 4(c)), a great majority of the respondents (over 67.0%) had positive attitudes towards the platform being easy to use, learn and navigate.

---

[21]http://inct.cnpq.br, accessed in February 2021.

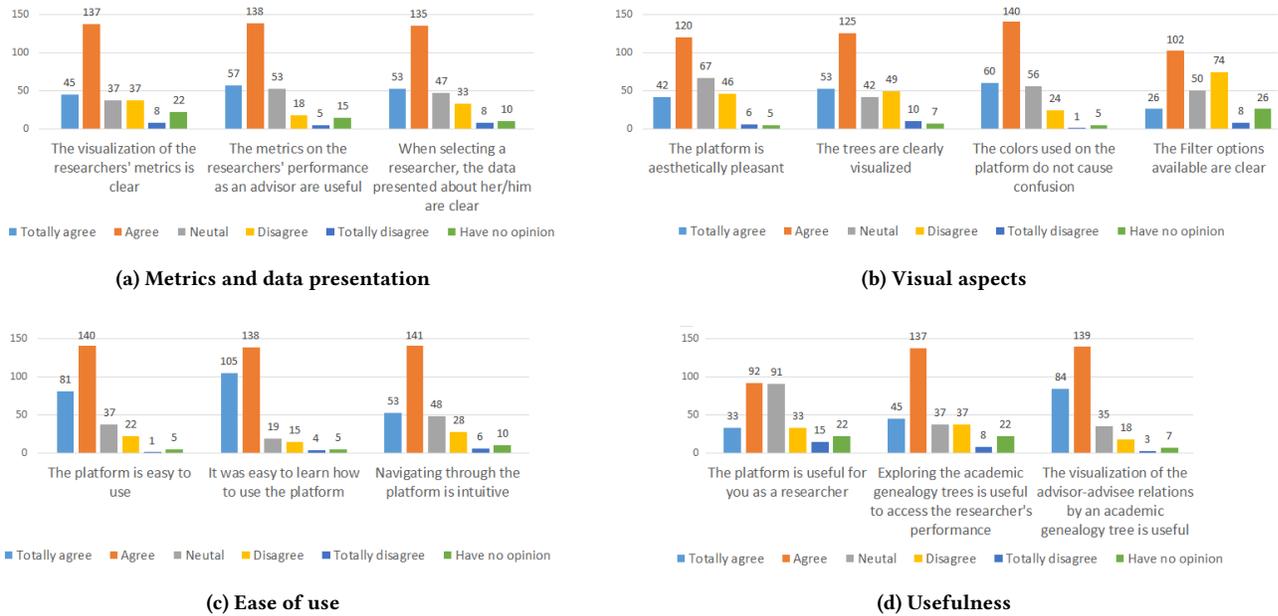

Figure 5: Respondents' opinion regarding specific aspects of the *Science Tree* platform.

Finally, regarding *usefulness* (Figure 4(d)), again a great majority of the respondents (over 75.0%) agreed that exploring the researchers' academic genealogy trees is useful to assess their performance, and that visualizing the advisor-advisee relationships as a tree is also useful as a form of expressing this academic role. However, their opinion about how useful the platform itself would be to them as researchers was not so positive, i.e., 43.7% agreed it would be useful, 39.5% were neutral or did not have an opinion, and 16.7% did not agree. Overall, these responses seem to indicate that visualizing the academic genealogy trees is useful as an assessment task, but not so much from a research viewpoint.

***Qualitative Analysis.*** The results addressed here are based on a qualitative analysis of both the open-ended comments of the questionnaire and user evaluation sessions carried out with the seven prospective users. The analysis was conducted in three steps. First, we performed an inductive analysis (or bottom-up approach) [5] of the comments obtained from the survey, identifying the main themes or categories that emerged from them. Then, we analyzed the users' interviews in order to identify the main problems and issues raised by them. Finally, we used the categories generated in the previous step to organize and classify the main issues raised by the users.

---
[22]These filter options aimed at facilitating the search on the platform.

The inductive analysis of the 132 comments collected allowed us to classify them in four main categories: (i) *problems experienced with the platform*, (ii) *possible improvements*, (iii) *suggestions of new aspects to consider* and (iv) *usefulness*. These categories were then used to organize the issues identified in the users' evaluation[23]. It is important to note that the interview analysis provided a deeper understanding of the users' perspectives, thus allowing us to present the final results of this analysis organized by the above four categories.

Regarding *problems experienced with the platform*, they included difficulties faced by the users while using the platform, such as problems with different devices (e.g., smartphones and tablets) or browsers, and those related to the visualization of the trees. In the comments provided in the questionnaire, the respondents indicated that large trees are difficult to visualize, and in the interview participants not only pointed to the problem, but also explained the main reasons for this difficulty – the overlapping of the edges and the size of the nodes. *Possible improvements* included aspects that the users felt could enhance the platform usability or their experience, such as allowing for an (easier) way to explore ancestors in the trees[24], or improving the visualization of large trees, i.e., when a researcher has a large number of advisees. They also mentioned that clearer instructions and explanations about some of the functionality and interface elements would be useful.

---
[23]During this step of the analysis, the researchers observed if there were issues that could not be classified according to the four categories, but none were identified
[24]Exploring ancestors was possible, but many users did not identify how to do it, and suggested it should be included.

With respect to *suggestions of new aspects to consider*, they were of two different types. The first type was related to the visualization of the genealogy trees, whereas the second one aimed at broadening the scope of the platform to take into consideration other specific analyses. The suggestions regarding the genealogy trees included taking into account other types of advisory, such as undergraduate projects, or providing more specific information about the advisors (e.g., the institution from which they obtained their PhD dgrees) or advisees (e.g., their current affiliations). The broader suggestions went beyond the academic genealogy and considered other aspects related to the researchers' scientific performance, such as their co-authorship networks or specific issues about their publications.

Finally, *usefulness* referred to participants' comments on how useful or not they perceived the platform to be. Some of the respondents mentioned in the questionnaire that the visualization provided by the platform was interesting, but they did not imagine they would actually use it in their everyday activities. On the other hand, some others commented that they believed it would be useful. One use mentioned as relevant for researchers is to analyze their own productivity, based not only on the students they advised, but also the unfolding results from them. Others mentioned that it could be useful to support more administrative aspects, such as the impact of a graduate program, by analyzing the performance of its faculty members.

**Concluding Remarks.** Our results reveal that our quantitative and qualitative results are aligned. Regarding the limitations of the *Science Tree* platform, both analyses highlighted that the visualization of the academic genealogy trees needs some improvements. Other limitations highlighted by the users were the absence of detailed explanations about the metrics and the low visibility of the filters that ended up almost unused. On the other hand, the most notable positive point was the ease of use of the platform, recognized by 85.2% of the respondents of the questionnaire and by five out of the seven interviewees. Another highlight was the data presentation, with an approval rate of 65.3% by the respondents of the survey and by four out of the seven interviewees. In addition, overall, the interviews did not indicate new issues that had not been identified through the questionnaire, but did provide an in-depth understanding of the reasons or motivations for some of the results.

The conclusion obtained from these two user evaluations is that, although there is room for improvements, the *Science Tree* platform is easy to use and presents itself as a relevant tool to analyze the formation of researchers in Brazil. It is also important to note that some of the limitations raised by the end users during the platform evaluation have already been resolved, such as the inclusion of a page explaining the metrics provided about the genealogy trees and improvements on its help messages. However, there are still some specific aspects that need to be addressed as, for example, the visualization of trees with many levels and a large number of nodes, and the positioning of certain elements on the screen, such as the filters used for better defining the scope of the queries.

## 6 CONCLUSIONS AND FUTURE WORK

In recent years, there has been a great effort to develop specific platforms and applications aimed at building and maintaining academic genealogy trees of researchers from distinct knowledge areas [9, 10, 14, 19, 24]. Most of these initiatives adopt a collective collaboration strategy, in which such platforms or applications grow with data inserted by several collaborators. In this paper, we reported an effort carried out to develop a platform for building and visualizing the academic genealogy of Brazilian researchers. For this, we used data extracted from the curricula vitae of researchers with a PhD degree, collected from the CNPq's Lattes platform in XML format. Unlike other similar initiatives, such as the *Mathematics Genealogy Project*, which are limited to a single knowledge area, our effort allows the construction of academic genealogy trees of researchers from distinct areas.

Moreover, the *Science Tree* platform does not only show the researchers' academic genealogy trees, but also allows its users to explore them as they may be expanded with the nodes of their academic descendants. The platform also provides specific metrics related to the generated trees and the corresponding researchers, in addition to displaying data of interest extracted from the researchers' Lattes curricula. Thus, by exploring the *Science Tree* platform, it is possible to visualize the role of such researchers in the formation of the Brazilian academic community, many of them, such as Prof. Crodowaldo Pavan whose academic genealogy tree is presented in Section 3, are pioneers in their respective research areas and, therefore, played a leading role in the formation of important generations of Brazilian researchers. The use of the framework *.Net Core* with the *ReactJS* and *Vis.JS* visualization libraries, together with the *MariaDB* DBMS, provided the functionality required by the platform with good performance. In addition, its interface has shown a positive impact on the users' interaction experience, mainly due to the use of modern techniques for developing web applications, which provided the adequate facilities for data visualization.

To assess its facilities, the platform was evaluated by two groups of users, the first one consisting of 286 researchers who answered an evaluation questionnaire with 13 multiple-choice questions, and the second one, composed of seven researchers with large experience as graduate advisors, who agreed to participate in a face-to-face assessment conducted through a personal interview, during which they had to perform some tasks using the platform.

The results of these two evaluations with typical users enabled us not only to validate the main facilities provided by the platform, but also to identify design decisions that could be improved and new features that could be added to it in the future. Moreover, some of the points raised have already been taken into account, such as more detailed explanations about the current metrics provided for assessing the trees, but others will be considered as future improvements, like the inclusion of new evaluation metrics. The visualization of large trees is also a point that needs some improvements to provide other means to explore all its levels. It is also worth noting that the *Science Tree* platform is already available for wide use, and although it has not been widely publicized to the Brazilian scientific community yet, it has already been spontaneously accessed by more than 800 users from 25 out of 26 states in Brazil and also from 10 other countries according to data from Google Analytics[25].

When compared with similar initiatives (see Table 1), the *Science Tree* platform shows some characteristics that make the experience of exploring its contents very special, even considering the fact that it addresses the research community of a specific country. Its graphical interface is very intuitive and provides its users with an attractive view of the Brazilian academic genealogy. The metrics provided by the platform, although outnumbered by those available in *The Academic Family Tree*, are able to provide a clear view of the researchers' role as advisors. When compared with the *Acacia Platform* and *The Gold Tree*, the other two platforms that rely on data extracted from the Lattes platform, its graphical representation of the Brazilian academic genealogy provides a more intuitive view of how specific research groups have been created and grew up. Besides, the fact that personal data about the researchers are also directly extracted from the Lattes platform, and not individually provided by them, also makes the *Science Tree* platform much richer with respect to the data it provides to its users.

With respect to the architecture of the *Science Tree* platform, a major improvement required is to provide an automatic update of its data repository by periodically crawling the CNPq Lattes platform, thus making its content even more attractive to the Brazilian scientific community. Another point to be considered is to improve the algorithm currently used to generate the academic genealogy trees [14] to allow a more effective disambiguation of the researchers with similar names or whose names appear with different spellings in the Lattes platform, which can be carried out by using some of the methods previously developed by our research group [17]. Finally, it is worth emphasizing the importance of expanding the platform's dissemination in order to increase its use and the demand for new features.

## ACKNOWLEDGMENTS

This work has been partially supported by grants from FAPEMIG (grant APQ-02302-17) and CNPq (grant 308528/2019-0). The authors are very grateful to all researchers that participated in the end user evaluation reported in this article.

---

[25]https://analytics.google.com, accessed in February 2021.